\documentclass[aip,jcp,floatfix,reprint]{revtex4-1}

\usepackage{graphicx}% Include figure files
\usepackage{dcolumn}% Align table columns on decimal point
\usepackage{bm}% bold math
\usepackage{epsfig}
\usepackage{amsmath}
\usepackage{amssymb}
\usepackage{xspace}

\begin{document}

\title{Layering Transitions and Solvation Forces in an Asymmetrically Confined Fluid}
\author{M C Stewart}
\author{R Evans}
\email{bob.evans@bristol.ac.uk}
\affiliation{H. H. Wills Physics Laboratory, University of Bristol, Bristol BS8 1TL, United Kingdom}%

\date{\today}

\begin{abstract}
We consider a simple fluid confined between two parallel walls (substrates), separated by a distance $L$. The walls exert competing surface fields so that one wall is attractive and may be completely wet by liquid (it is solvophilic) while the other is solvophobic. Such asymmetric confinement is sometimes termed a `Janus Interface'. The second wall is: (i) purely repulsive and therefore completely dry (contact angle $\theta=180^{\circ}$) or (ii) weakly attractive and partially dry ($\theta$ is typically in the range $160-170^{\circ})$. At low temperatures, but above the bulk triple point, we find using classical density functional theory (DFT) that the fluid is highly structured in the liquid part of the density profile. In case (i) a sequence of layering transitions occurs: as $L$ is increased at fixed chemical potential $\mu$ close to bulk gas--liquid coexistence $\mu_{co}$, new layers of liquid-like density develop discontinuously. In contrast to confinement between identical walls, the solvation force is repulsive for all wall separations and jumps discontinuously at each layering transition and the excess grand potential exhibits many metastable minima as a function of the adsorption. For a fixed temperature $T=0.56T_C$, where $T_C$ is the bulk critical temperature, we determine the transition lines in the $L$, $\mu$ plane. In case (ii) we do not find layering transitions and the solvation force oscillates about zero. We discuss how our mean-field DFT results might be altered by including effects of fluctuations and comment on how the phenomenology we have revealed might be relevant for experimental and simulation studies of water confined between hydrophilic and hydrophobic substrates, emphasizing it is important to distinguish between cases (i) and (ii).
\end{abstract}

%\pacs{68.08.Bc,05.70.Np,68.15.+e}% PACS, the Physics and Astronomy
                             % Classification Scheme.%
%\keywords{Adsorption, Density Functional Theory, Confined fluids, Interface Structure}%Use showkeys class option if keyword
                              %display desired
\maketitle

\section{Introduction}
\label{sec:int}

Understanding how confinement influences the static and dynamic properties of fluids has relevance for several different areas of physical chemistry and for materials science and engineering. For example, ascertaining the nature of adsorption at solid adsorbants and in nano and mesoporous media is of practical interest in many industrial processes \cite{Dab}. Similarly understanding static and dynamic wetting phenomena on the nanoscale is important for developments in nanofluidics and possibly in nanobiotechnology \cite{RauDie}. Fundamental studies of the effects of confinement have naturally focused on simple confining geometries. Thus much is now known about the  phase behaviour of fluids confined  in symmetric geometries such as between two identical planar walls or in a cylindrical pore \cite{GelbGub,Evans90,BinLanMul,BinHorVinVir,MulBin}: confinement causes shifts of phase boundaries from those in bulk and can give rise to entirely new phases which have no bulk fluid analogue. Far less is understood about fluids confined asymmetrically. The properties of such a system are of interest, for example, in microfluidics, in the self-assembly of molecules and in nanolubrication. In biology, protein molecules are comprised of both hydrophilic and hydrophobic sites so an understanding of the forces arising from confining a fluid between substrates with different (competing) adsorbing properties may play a role in determining how proteins fold \cite{HuangChandP,Chand05}.

Layering arises from packing effects in fluids adsorbed at walls. The best studied situation is that of gas adsorption at a single attractive wall. Enhanced attraction between the fluid and a solvophilic wall can result in a region of fluid near to the wall that is very much higher in average density than the bulk fluid (a gas) and layering transitions can occur as, one by one, new high density layers are adsorbed at the solvophilic wall as the chemical potential $\mu$ is increased towards bulk gas--liquid coexistence $\mu_{co}$ at a fixed (low) temperature. In this paper we show that layering transitions may also occur on increasing the separation, $L$, between two parallel walls that exert competing solvophilic and solvophobic confining surface fields.  The first-order layering transitions are seen as jumps in the adsorption isotherm as $L$ increases, at fixed $\mu$ close to $\mu_{co}$, and there is sufficient space for another adsorbed liquid layer to develop.

Layering transitions were discovered experimentally using volumetric and gravimetric adsorption techniques. Sharp steps in the adsorption isotherms at low temperatures for various gases adsorbed on graphite indicated the possibility of a series of first-order phase transitions \cite{ThomDuvReg}. Exfoliated graphite has proved a particularly suitable substrate due to its large, homogeneous surface area. More sophisticated techniques such as ellipsometry, neutron diffraction and x-ray scattering have extended the number of layering transitions observed, e.g., eight for ethylene on graphite \cite{DrirNhamHess}, nine for oxygen \cite{DrirHess} and ten for nitrogen \cite{VolkKnor} and have allowed the identification of phase transitions within the adsorbed layers \cite{ZhuDash,ZhuDashB,LarZha,YounMengHess}. Some of the early experiments have been reviewed by Thomy {\it{et al}} \cite{ThomDuvReg} and later summarised \cite{ThomDuv}. An important issue is whether or not the discrete (first-order) jumps in adsorption correspond to transitions between liquid, as opposed to solid, adsorbed phases; in many cases the transitions persist to temperatures above the bulk triple point. A theoretical perspective on layering transitions is given in Refs. \citenum{SulTdG,EvR,GelbGub}. Early theoretical studies of layering transitions focused on lattice-gas models for the fluid \cite{OlivGrif,Eb80,Eb81,PanSchWor}. Pandit {\it{et al}} \cite{PanSchWor} performed a systematic investigation of adsorption on attractive substrates and found first order layering transitions for sufficiently strongly attractive substrates. The transitions terminate in (two-dimensional Ising like) critical points at temperature $T_{cn}$ for the $n$th transition and as $n\rightarrow\infty$, $T_{cn}$ approaches the roughening temperature $T_R$ of the lattice-gas. For strongly attractive substrates the wetting transition temperature $T_W$  lies below $T_R$  and wetting occurs via an infinite sequence of layering transitions as bulk liquid--gas coexistence is approached. For weaker substrates the layering transitions coalesce into a single thin-film to thick-film (prewetting) transition whose critical point lies slightly out of bulk coexistence, above but close to $T_W$. However, the imposed discrete structure of the lattice-gas model overemphasizes the layer-like structure of adsorbed films. Because of the particle-hole symmetry, layering transitions are even seen in drying films, i.e. in adsorption from a bulk liquid at a repulsive or a very weakly attractive substrate; this is unrealistic \cite{PanSchWor}. The physical relevance of the lattice-gas model is restricted to the situation where dense films are adsorbed such that hard-core interparticle repulsion leads to the formation of liquid-like layers.

Continuum, off-lattice models are clearly more relevant for real fluids, and layering transitions have been investigated using classical density-functional theories (DFT) and in simulation. In a pioneering DFT study Ebner and Saam \cite{EbSaam77} found, using a non-local free-energy functional for a model of argon at a weakly adsorbing solid ${\mathrm {CO_2}}$ substrate, a single first-order transition from an adsorbed film of thickness one or two (atomic) layers to one of several layers. This transition was later identified as an example of Cahn's generic prewetting.

Ball and Evans \cite{BallEv} employed a weighted density approximation for the hard-sphere part of the Helmholtz free-energy functional to investigate simple model fluids adsorbed at strongly attractive structureless (planar) walls. They found a large number of layering transitions (nine or ten) at temperatures between about $0.5T_C$ and $0.6T_C$, where $T_C$ is the bulk critical temperature. Subsequently Fan and Monson \cite{FanMon} investigated a system in which the fluid molecules interact with a shifted force Lennard-Jones (LJ) 12-6 potential and are adsorbed at a weakly attractive, planar 9-3 wall. For this system they found prewetting followed by a sequence of layering transitions as $\mu$ was increased towards $\mu_{co}$ at temperatures above the bulk triple point temperature of the fluid.  These authors\cite{FanMon} found good qualitative agreement between the adsorption isotherms obtained from their Monte Carlo simulations and those they obtained using a version of DFT similar to that of Ball and Evans \cite{BallEv}.

More recently layering transitions were found in DFT studies of the Asakura-Oosawa (AO) model of colloid-polymer mixtures adsorbed at a planar hard wall  \cite{BradEvSchLow,BradEvSch,WesSchLow}. As a result of depletion attraction the hard wall favours the phase (liquid) that is rich in colloid and transitions occur on adsorbing from the bulk colloid-poor phase (gas). The polymer reservoir packing fraction, $\eta_p^r$, plays a role equivalent to inverse temperature.  Depending on the colloid-polymer size ratio, three or four layering transitions followed by what appears to be a first-order wetting transition were found as $\eta_p^r$ decreased along bulk coexistence. Unlike the layering transitions observed in simple liquids at attractive walls, these entropically driven transitions \cite{BradEvSch} occur far from the bulk triple point. Monte Carlo simulations \cite{DijVR} for the same AO model also find layering transitions and although the phase diagram is not identical to that obtained from DFT calculations \cite{BradEvSch} similar features emerge. Moreover, the density profiles of colloid and polymer species found in simulation are close to those from DFT.

When a fluid is confined between two identical walls or inside a cylindrical pore its phase behaviour may change markedly from in bulk. For solvophilic walls the bulk gas--liquid coexistence line is shifted to a lower chemical potential; this is termed capillary condensation. The same type of confinement also causes the prewetting and layering transition lines found for the single wall to shift to different locations in the $\mu$, $T$ phase diagram and to compete with capillary condensation \cite{Evans90}. Ball and Evans \cite{BallEv} used DFT to study layering of model fluids confined in cylindrical pores and found that the layering transitions occur at lower pressures compared to the single wall system.  The sequence of transitions between stable layered states was truncated by the onset of capillary condensation, i.e. the pore was filled by liquid at a certain chemical potential and for larger $\mu$ the layering transitions were metastable. Grand canonical Monte Carlo (GCMC) simulations of a L-J fluid in cylindrical pores \cite{PetHefGubSwol} found similar results. Layering transitions were also observed for a simple model of methane adsorbed in symmetric slit-like pores using GCMC and Molecular Dynamics (MD) simulations \cite{JiaRhyGub}; as the wall separation  $L$ increased a greater number of stable  layering transitions was found since then capillary condensation occurs closer to $\mu_{co}$ \cite{Evans90}.

In this paper we investigate layering in a L-J like fluid adsorbed in a slit-like pore where the two confining walls are competing, i.e. one wall is solvophilic and the other is solvophobic. The work of Parry and Evans  \cite{ParryEvPRL, ParryEv} established that the phase behaviour of fluids under such confinement can be very different from that found when the two walls are identical. For the case where the solvophobic wall is dry (contact angle $\theta=180^{\circ}$), phase coexistence is suppressed until temperatures below $T_W$, the wetting transition temperature of the solvophilic wall. In the temperature range $T_C>T>T_W$  there is a single ‘soft-mode’ phase characterized by a density profile that has low density gas near to the solvophobic wall and high density liquid  near to the solvophilic wall. For large $L$ and $\mu$ near $\mu_{co}$ there is a strongly fluctuating delocalized gas--liquid interface\cite{ParryEv,VirVinHorBin,StewEv12} centered near $L/2$ . In a recent paper \cite{StewEv12} we performed a detailed DFT investigation of the `soft mode' or `delocalized interface' phase of a model fluid similar to the one we consider here. That study focused on a single temperature $T=0.8 T_C>T_W$ and very large values of $L$ so that we could investigate the predictions of an effective interfacial Hamiltonian approach for the scaling properties of thermodynamic quantities in the ‘soft-mode’ phase of a model fluid subject to dispersion forces; we deliberately retained the long-range ($-r^{-6}$) tails of the interatomic potentials. 

The thrust of our present study is very different. We concentrate on lower temperatures, close to the bulk triple point $T_{tr}$ and separations $L$ typically $<20$ atomic diameters. Two types of solvophobic wall are considered: (i) the wall--fluid potential is purely repulsive, as in Ref.\ \citenum{StewEv12}, and therefore this wall is always dry and (ii) the wall--fluid potential has a weak attractive piece so that the wall is partially dry, i.e., we choose parameters in order that the contact angle $\theta$ is large, in the range $160^{\circ}-170^{\circ}$.  We find that the phase behaviour and properties of the confined fluid differ dramatically between cases (i) and (ii).

One of the motivations for our present study was to ascertain how $f_s$, the solvation force, i.e. the excess pressure or force per unit area between  two walls arising from confinement of the fluid, differs from the well-studied case of identical walls. In the latter case $f_s$ is usually attractive at large $L$, although the precise form can depend on the details of the wall--fluid and fluid--fluid potentials and the thermodynamic state point of the fluid \cite{MacDrzBry}. Determining the form of $f_s$ is important generally for understanding solvent-mediated interactions, especially in colloidal systems. For example the aggregation of colloidal particles depends on the solvent-mediated effective potential  between two big colloids which can be obtained from $f_s$ using the Derjaquin approximation \cite{Der1934}. We note that much current research is concerned with the near-critical regime of the solvent where the solvation force acquires a so-called critical Casimir contribution which has a universal scaling form that depends on the proximity to the solvent critical point, expressed in terms of the ratio of $L$ to the diverging bulk correlation length, and on the nature of the confining walls. For identical walls the scaling function for the critical Casimir force is attractive while for competitive (opposing) walls it is repulsive.

The development of the surface force apparatus (SFA) by Israelachvili and co-workers \cite{Isr} and of atomic force microscopy (AFM) techniques led to many experimental investigations of $f_s$; see for example the review \cite{ChrisClaes}. A notable early SFA experiment by Horn and Israelachvili \cite{HorIsr} in 1980 on an inorganic liquid between mica cylinders found oscillatory behaviour with the maxima of $f_s$ decaying exponentially with increasing separation. One year earlier Lane and Spurling \cite{LanSpu} and van Megen and Snook \cite{MegSno} reported oscillatory $f_s$ in MC simulations of a LJ fluid confined between identical planar (solvophilic) walls. Since these pioneering investigations there have been many studies of oscillatory $f_s$ and it is clear that the oscillations arise from the packing of the atoms in the confined fluid, i.e. the same physics that is responsible for the layered structure of the density profile close to the walls. Note that some evidence for layering transitions has been found in AFM 
measurements of the shear response of a confined fluid, e.g., for water between a glass tipped probe and a mica surface \cite{AntHumMil}. Much less is known about the solvation force under asymmetric confinement, a subject of our present study.

An experiment by Zhang {\em et. al.} \cite{ZhaZhuGran} measured, in the SFA, the response to shear deformations of water confined between a hydrophilic and hydrophobic surface (a Janus interface). Giant fluctuations in the viscous response were seen which the authors \cite{ZhaZhuGran} suggested were due to the presence of a wandering liquid--gas interface between the two surfaces of the type predicted theoretically by Parry and Evans \cite{ParryEvPRL, ParryEv}. Motivated by the experimental results, McCormick \cite{McC} performed lattice gas simulations of `water' between a hydrophobic and a hydrophilic plate. He found large fluctuations in the contact density at the hydrophobic plate which were attributed to a fluctuating liquid--gas interface. Subsequent grand canonical Monte Carlo simulations by Pertsin and Grunze \cite{PertGrun} for TIP4P water showed large fluctuations in the number of `water' molecules present between hydrophilic and hydrophobic walls and the form of the density profiles indicated the presence of a wandering liquid--gas interface for the case of a purely repulsive hydrophobic wall. These studies provided some of the motivation for our current investigation.

In this paper we first study the adsorption behaviour of a model fluid, described in Section \ref{sec:Model}, at a single solvophilic wall, over a range of temperatures from below the bulk triple point ($T_{tr}\approx 0.5T_C$) up to $0.65T_C$. We have identified the equilibrium layering transitions for $\mu<\mu_{co}$ and at coexistence by comparing the grand potential of states with different numbers of layers at each state point. Unlike previous DFT studies we have traced accurately the locations of the first three layering transition lines up to their critical points and have produced a phase diagram in the $(T,\mu)$ plane. Focusing on the $T=0.56T_C$ isotherm we show examples of density profiles at the layering transitions and plot the adsorption as bulk liquid--gas coexistence is approached. Our DFT method is described in Section \ref{sec:DFT} and results are given in Section \ref{sec:Semi-inf}. 

In Section \ref{sec:confined_results} we confine our model fluid between two competing walls with sufficiently small wall separations that the, purely repulsive, solvophobic wall (system (i)) influences the layers of liquid next to the solvophilic wall.  We find that the chemical potential at which the layering transitions occur depends on the wall separation $L$ and we have determined the full phase diagram as a function of chemical potential $\mu$ and wall separation $L$ at one temperature $T=0.56T_C$.  In this system there is no capillary condensation as bulk liquid--gas coexistence is approached from the gas side with the interesting consequence that layering transitions can occur at bulk coexistence and even on the liquid side $\mu>\mu_{co}$.

We investigate (Sec.\ \ref{sec:fs}) the variation of the solvation force $f_s$ with wall separation $L$ for a liquid confined between identical, solvophilic walls, comparing and contrasting the results with those for $f_s$ obtained for the same model system confined in two different asymmetric slits. Firstly a purely repulsive wall--fluid interaction (i) was chosen so that the solvophobic wall is completely dry in the semi-infinite system and secondly we consider a weakly attractive wall--fluid potential (ii) so that the wall is solvophobic but does not exhibit complete drying. The same solvophilic wall potential was used in all three systems. In Section \ref{sec:PG} we compare density profiles for the same model system obtained using our DFT with those from the simulations of asymmetrically confined ‘water’ performed by Pertsin and Grunze \cite{PertGrun}. We find striking similarities, including the presence of metastable states.

 In Section \ref{sec:Dis} we summarise our results and discuss the limitations of DFT, including the possibility of the layers freezing and the effects of thermal capillary wave fluctuations. We compare results for the solvation force with those of a previous DFT study by Balbuena {\it et. al.}\cite{BalBerGub} and comment on their relevance for the experiments of Zhang {\it et. al.} \cite{ZhaZhuGran} and the lattice gas simulations of McCormick \cite{McC}.

\section{The Model and the DFT Approach}
\label{sec:Model}
Our model system was deliberately chosen to be similar to that investigated by Pertsin and Grunze \cite{PertGrun}.
\subsection{Fluid--fluid potential}
\label{sec:FFP}
In a study by Pertsin and Grunze \cite{PertGrun} the well known TIP4P model  \cite{JorgChand} was used in simulations of `water' molecules. Our DFT model is much simpler than this---the molecules are approximated by hard-spheres with an attractive cut and shifted Lennard-Jones potential between the centres of the spheres:
\begin{equation}
\phi_{att}(r)=
\begin{cases}
-\epsilon - 4\epsilon\left[\left(\frac{\sigma}{r_c}\right)^{12}-\left(\frac{\sigma}{r_c}\right)^6\right] \hspace{5mm} r<r_{min}
\\
4\epsilon\left(\left[\left(\frac{\sigma}{r}\right)^{12}-\left(\frac{\sigma}{r}\right)^6\right]- \left[\left(\frac{\sigma}{r_c}\right)^{12}-\left(\frac{\sigma}{r_c}\right)^6\right]\right)
\\
 \hspace{39mm}  r_{min}<r<r_c
\\
0 \hspace{49mm}  r>r_c
\end{cases}
\label{eq:ffpot}
\end{equation}
where $\sigma=3.15$\AA, $r_{min}=2^{1/6}\sigma$, the cutoff $r_c=2.2831\sigma=7.2$\AA\ and $\epsilon=1.55$ kcal/mol ($\epsilon/k_B=780$K). This attractive potential is close to the potential between pairs of oxygen atoms in the `water' molecules in the TIP4P model used in simulations. The hard-sphere diameter is taken to be $d=3.00$\AA\ $=0.95\sigma$. Unlike the TIP4P model there are no Coulomb interactions between our fluid molecules so the net short-range attraction between molecules is much less than that in simulations. Obviously our model will not exhibit any effects of hydrogen-bonding; it is certainly a zeroth-order, orientation independent model of water.

\subsection{Wall--fluid potentials}
\label{sec:WFP}

The wall potentials were chosen to be close to the potentials between the wall and the water oxygen atoms in the unstructured (planar) walls used by Pertsin and Grunze \cite{PertGrun}. The solvophilic wall, positioned at $z=0$, is a cut and shifted Lennard-Jones ($9,3$) potential:
\begin{equation}
V_w(z)=
\begin{cases}
\frac{\epsilon_w}{2}\left[\left(\frac{\zeta}{z}\right)^9-3\left(\frac{\zeta}{z}\right)^3-\left(\frac{\zeta}{z_c}\right)^9+3\left(\frac{\zeta}{z_c}\right)^3\right]  z<z_c
\\
0 \hspace{8mm} z>z_c,
\end{cases}
\label{eq:wpot}
\end{equation}
where $\zeta$ is a measure of the range of the potential. We set $\zeta=d$. The cutoff is the same as for the fluid--fluid potential, $z_c=r_c$, and the strength of the potential is $\epsilon_w=6.953$ kcal/mol (giving a well depth of $6.2$ kcal/mol). This potential is {\em not} the same as that obtained by integrating a potential of the form of Eq.\ (\ref{eq:ffpot}) over a semi-infinite volume, i.e., the wall is not equivalent to a volume of the fluid at constant density.

When we investigate the asymmetrically confined fluid, we position a solvophobic wall at $z=L$, parallel to the solvophilic wall. We consider two different wall--fluid potentials for the solvophobic wall:
\begin{description}
\item[(i)] A purely repulsive potential,
\begin{equation}
V_w^{rep}(z)=
\begin{cases}
\frac{\epsilon_w}{2}\left[\left(\frac{\zeta}{z}\right)^9-3\left(\frac{\zeta}{z}\right)^3-\left(\frac{\zeta}{z_c}\right)^9+3\left(\frac{\zeta}{z_c}\right)^3\right] z<z_0
\\
0 \hspace{8mm} z>z_0,
\end{cases}
\label{eq:wpotr}
\end{equation}
where $z_0$ is the point where the wall potential in Eq.\ (\ref{eq:wpot}) crosses the $z$-axis, i.e. $V_w(z_0)=0$. The strength of the potential is $\epsilon_w=0.516$ kcal/mol.
\item[(ii)] A weakly attractive wall, exerting a Lennard-Jones ($9,3$) potential with the same form as the solvophilic wall (\ref{eq:wpot}), but with $\epsilon_w=0.516$ kcal/mol, giving a well depth of $0.46$ kcal/mol, a factor of $13.5$ smaller than the solvophilic wall --- see Fig.\ 1 of Pertsin and Grunze \cite{PertGrun}. This is the same potential as that used in the simulations \cite{PertGrun}, where the parameters were chosen to describe the interaction of a water molecule with a hydrophobic paraffin surface.
\end{description}

\subsection{Density Functional Theory approach}
\label{sec:DFT}

In density functional theory (DFT) the free energy of an inhomogeneous fluid is expressed as a functional of the average one-body density $\rho(\bf{r})$ (for a review of DFT see Ref.\ \citenum{EvR}). In the approximation that we employ, the excess hard sphere part of the Helmholtz free energy functional ${\cal F}_{ex}^{hs}$ is treated by means of Rosenfeld's fundamental measures theory \cite{Rosen89} and the attractive part of the fluid--fluid interaction potential is treated in mean-field fashion. This free-energy functional has the same form as that used in Refs.\ \citenum{StewEv12,StewEv}, where it is described in more detail. The grand potential functional is
\begin{eqnarray}
\Omega_V[\rho]&=&{\cal F}_{id}[\rho]+{\cal F}_{ex}^{hs}[\rho]  \nonumber
\\
&& \mbox{}+\frac{1}{2}\int{\int{{\rm d}{\mathbf r}_1{\rm d}{\mathbf r}_2\ \rho({\mathbf r}_1)\rho({\mathbf r}_2)\phi_{att}(\lvert{\mathbf r}_1-{\mathbf r}_2\rvert)}} \nonumber
\\
&&\mbox{}+\int{\rho({\mathbf r})(V({\mathbf r})-\mu)\,{\rm d}{\mathbf r}},
\label{eq:FGP}
\end{eqnarray}
where the density profile $\rho({\mathbf r})=\rho(z)$. The external potential corresponding to the semi-infinite fluid at a single solvophilic wall is $V({\mathbf r})\equiv V_w(z)$ with $V_w(z)$ given by Eq.\  (\ref{eq:wpot}). For the fluid confined between two parallel walls $V({\mathbf r})\equiv V(z;L)=V_{w1}(z)+V_{w2}(L-z)$, where the solvophilic wall 1 is at $z=0$, with $V_{w1}$ given by Eq.\  (\ref{eq:wpot}). The solvophobic wall 2 is at $z=L$, and $V_{w2}$ is given by Eq.\ (\ref{eq:wpotr}) for system (i), where wall 2 is purely repulsive, and by Eq.\  (\ref{eq:wpot}) for system (ii) in which wall 2 is weakly attractive. ${\cal F}_{id}[\rho]$ is the Helmholtz free energy functional for the ideal gas. The attractive fluid--fluid potential $\phi_{att}$ is given by Eq.\  (\ref{eq:ffpot}), and with this choice the homogeneous fluid described by Eq.\ (\ref{eq:FGP}) has a (mean-field) critical temperature $k_BT_C/\epsilon=1.35$ and density $\rho_Cd^3=0.2457$. We did not investigate freezing within the DFT approach so we do not know the bulk triple point of the theory. However, we note that simulations \cite{AhmSad} for the full LJ pair potential give the ratio of triple point to critical point temperatures as $T_{tr}/T_C \lesssim0.52$. For the LJ potential truncated and shifted at $r_c=2.5\sigma$ the same ratio\cite{FanMon} is $< 0.55$. 

The equilibrium density profile was found by minimising the grand potential functional $\Omega_V[\rho]$. In Appendix \ref{sec:num} we describe the procedure we employed to deal with the highly oscillatory nature of the density profile in the vicinity of the solvophilic wall and the accompanying complication of having a large number of metastable states corresponding to different numbers of layers of liquid.

\section{Layering Transitions at a single Solvophilic Wall}
\label{sec:Semi-inf}
We began by investigating the layering transitions that occur in the semi-infinite fluid at the solvophilic substrate when the fluid is near to bulk liquid--gas coexistence, chemical potential $\mu\rightarrow\mu_{co}$. Initially the temperature was fixed at $T=0.56T_C$, which we estimate is just above the triple point temperature. At bulk coexistence $\mu_{co}^{-}$ the equilibrium state at this temperature, i.e., the density profile which minimises grand potential $\Omega_V[\rho]$, has seven layers of liquid next to the wall and there are many metastable states with both lesser and greater numbers of liquid layers.  Decreasing the chemical potential $\mu$ from its value at coexistence $\mu_{co}$, into the region of the phase diagram where the gas is the bulk stable state and the liquid is metastable, incurs a free energy cost proportional to the volume of liquid phase present and to the magnitude of the deviation from bulk coexistence $\delta\mu=\mu-\mu_{co}$. Since this free energy cost is greater for states with more liquid layers, the equilibrium state can change, as the chemical potential is decreased, to a state with fewer layers of the high density liquid and a layering transition can occur.
\begin{figure}
\centering
\epsfig{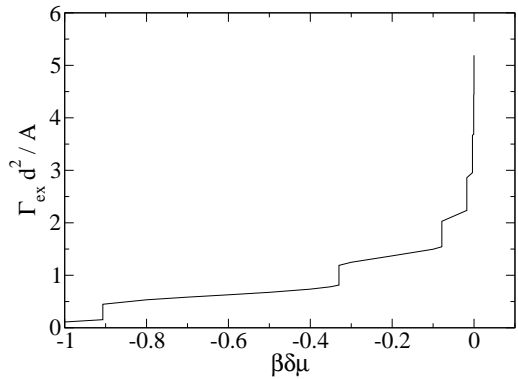}
\caption[Excess adsorption isotherm for the fluid at a single solvophilic wall at $T=0.56T_C$]{Excess adsorption isotherm for the fluid at a single solvophilic wall at $T=0.56T_C$. The metastable branches and spinodals are not shown. The first order layering transitions appear as steps in the adsorption isotherm. As the reduced chemical potential $\beta\delta\mu$ approaches bulk liquid--gas coexistence $\beta\delta\mu=0^-$, the transitions occur at smaller intervals. At coexistence there are seven adsorbed layers.}
\label{fig:Ads0_56}
\end{figure}
%\begin{figure}
%\centering
%\epsfig{figure=LGP0_56.eps}

%\caption[Surface Tension isotherm for $T=0.56T_C$]{Surface Tension isotherm for $T=0.56T_C$. The positions of the layering transitions are marked by $\bullet$.}
%\label{fig:GP0_56}
%\end{figure}

The adsorption isotherm at $T=0.56T_C$  is shown in Fig. \ref{fig:Ads0_56}. We define the Gibbs excess adsorption per unit area as $\Gamma_{ex}(T,\mu)/A=\int_0^{\infty}{dz[\rho(z)-\rho_b(T,\mu)]}$, where $\rho_b(T,\mu)$ is the bulk fluid density. The metastable branches, which are extensive, have not been plotted---only the adsorption for equilibrium states is shown. There are seven steps in the adsorption as bulk coexistence is approached, corresponding to first order layering transitions. The $7{\textsuperscript{th}}$ layering transition occurs at $\beta\delta\mu=-7.14\times10^{-5}$. The temperature is  below the wetting temperature $T_W$ because the adsorption $\Gamma_{ex}$ remains finite at bulk coexistence $\mu_{co}^-$. 

%Figure \ref{fig:GP0_56} gives the surface tension of the interface as a function of chemical potential deviation from coexistence at $T=0.56T_C$. The layering transitions are seen as kinks in the plot.  The surface tension increases with deviation from coexistence ($\beta\delta\mu=0^-$) as the reduction in free energy due to the presence of liquid rather than gas next to the wall competes with the increasing free energy cost resulting from the liquid phase no longer being stable in the bulk fluid.

\begin{figure}
\centering
\epsfig{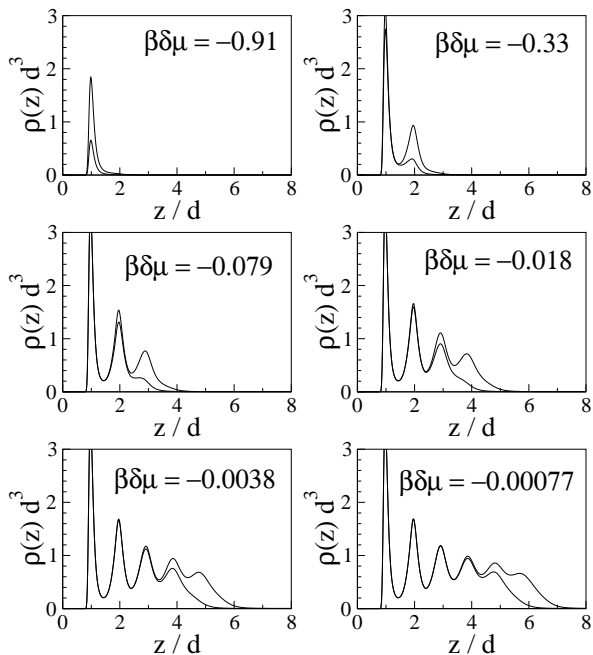}
\caption{Density profiles of the two coexisting states, with equal grand potential, at the first six layering transitions in the semi-infinite system consisting of the model fluid at the solvophilic wall, along the isotherm $T=0.56T_C$. These profiles correspond to the jumps in adsorption in Fig.\ \ref{fig:Ads0_56}. Note that there is also a 7th layering transition (not shown), which occurs at $\beta\delta\mu=-7.14\times10^{-5}$.}
\label{fig:sixprofs}
\end{figure}

Density profiles at each of the first six layering transitions along the isotherm $T=0.56T_C$ are shown in Fig. \ref{fig:sixprofs}. Away from bulk liquid--gas coexistence, before the first transition is reached, there is a thin layer of fluid next to the wall with higher density than the bulk gas. At the first layering transition, near $\beta\delta\mu=-0.91$, the fluid density in this layer increases discontinuously. At the second transition near $\beta\delta\mu=-0.33$ the jump in the adsorption arises from increases in the density of the fluid in both the second layer and the layer adjacent to the wall. Subsequent transitions show the formation of a new layer of denser fluid coupled with an increase in the density of the previous layer. The density profiles are highly oscillatory---even the fifth and sixth peaks can be discerned clearly. It is likely that the DFT exaggerates the layering structure i.e. the amplitude of the density oscillations and we return to this in Section \ref{sec:Dis}. The shapes of these density profiles are similar to those obtained from DFT by Ball and Evans \cite{BallEv} and by Fan and Monson \cite{FanMon} who considered similar model fluids and substrates.

\begin{figure}
\centering
\epsfig{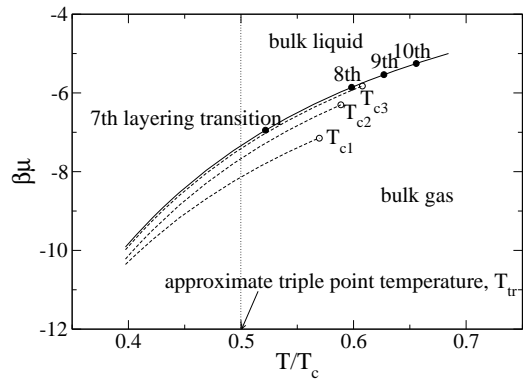}
\caption[Phase Diagram for the fluid at a single solvophilic wall at different temperatures $T$ and chemical potentials $\mu$]{Phase Diagram for the fluid at a single solvophilic wall at different temperatures $T$ and chemical potentials $\mu$. Bulk liquid--gas coexistence is marked with a solid line. Only the 1st, 2nd and 3rd transition lines (dashed lines) are shown.  Critical points $T_{cn}$ are shown as $\circ$.  For the 4th layer and above the transition lines lie very close to coexistence. The points where layering transitions first occur along bulk coexistence on increasing $T$ are marked with $\bullet$. The triple point temperature $T_{tr}$ is expected to be at about $0.5T_C$.}
\label{fig:PhaseD}
\end{figure}

Figure \ref{fig:PhaseD} displays the phase diagram in the $(T,\mu)$ plane for the semi-infinite fluid at the solvophilic wall. The locations of the 1st, 2nd and 3rd layering transitions are shown as dashed lines ending in critical points at temperatures $T_{cn}$. These are the {\em equilibrium} transitions, found by comparing the values of the minima in the excess grand potential $\Omega(\Gamma_{ex})$ as a function of the excess adsorption $\Gamma_{ex}$---see Appendix \ref{sec:num} for a description of our numerical methodology. Locating the transitions is not dependent on hysteresis effects. For the layering critical points shown it is clear that $T_{cn}<T_{c(n+1)}$ and this was also found to be the case for the higher order transitions, certainly up to $n=4$.  In Fig.\ \ref{fig:PhaseD} we mark the points where the 7th, 8th, 9th and 10th transitions first occur along bulk coexistence on increasing $T$. The 4th and higher order transitions lines occur very close to the bulk coexistence line and although we were able to trace their locations off bulk coexistence, these are not plotted in Fig. \ref{fig:PhaseD}. (Note that, although they lie close to coexistence, the lines are not short in temperature $T$. By contrast the 2nd and higher order transition lines calculated for a colloid-polymer mixture \cite{BradEvSch} are very short in polymer reservoir fraction $\eta_p^r$.) There are no layering transitions on the liquid side of coexistence. For $\mu>\mu_{co}$ the density profiles show oscillations near the wall which decay smoothly into the bulk liquid.

\section{Layering transitions in a fluid confined between a solvophilic and a solvophobic wall (`Janus Interface') }
\label{sec:confined_results}
\begin{figure}
\centering
\epsfig{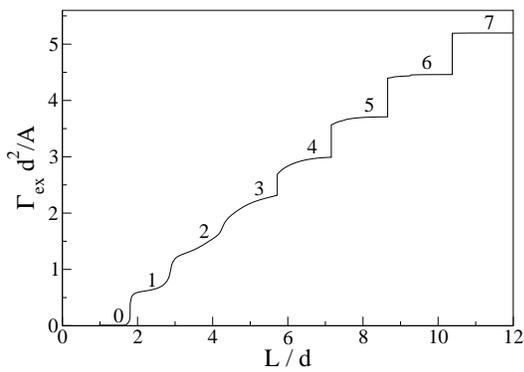}
\caption[Excess adsorption per unit area versus wall separation $L$ at bulk coexistence $\mu_{co}$ and $T=0.56T_C$ for system (i)]{Excess adsorption per unit area, $\Gamma_{ex}/A=\int_0^L dz (\rho(z)-\rho_b(T))$, versus wall separation $L$ at bulk coexistence $\mu_{co}$ and $T=0.56T_C$ for system (i)---the solvophobic wall is purely repulsive. The layering transitions 1-2 and 2-3 are continuous as $L$ is increased at this temperature. The other layering transitions are discontinuous (first order).}
\label{fig:PTWads}
\end{figure}

 In this section we describe DFT results for our model fluid confined between two parallel, planar walls: the solvophilic wall described in Sec.\ \ref{sec:Semi-inf} and a purely repulsive solvophobic wall [for details of the wall--fluid potential see system (i), Sec.\ \ref{sec:WFP}]. This wall is completely dry at all temperatures $T<T_C$ in the semi-infinite system consisting of the isolated wall in contact with the model fluid, i.e. the contact angle $\theta=\pi$. The temperature of our investigation is fixed at $T=0.56T_C$, which is sufficiently low that the layering transitions observed in the fluid at the isolated solvophilic wall on approaching bulk liquid--gas coexistence (see Sec.\ \ref{sec:Semi-inf}) are unlikely to have been washed out by fluctuations. Initially we fix the chemical potential of our system at bulk coexistence, $\mu=\mu_{co}$, and vary the wall separation $L$. Figure \ref{fig:PTWads} shows the adsorption isotherm. As $L$ is increased from zero the adsorption increases slowly until a jump at around $L=1.7d$ corresponding to the 1st layering transition. The development of the second and third layers appear as smooth increases in the adsorption. The 4th, 5th, 6th and 7th layering transitions are first order which is manifest in discrete steps in the adsorption isotherm. After the 7th layering transition at $L=10.38d$ there is little change in the adsorption as $L$ is increased because there are seven layers at the solvophilic wall in the semi-infinite system---any increase in $L$ beyond $10.38d$ in the confined system is simply accompanied by an increase in the amount of the low density gas phase causing no significant change to the excess adsorption.

\begin{figure}
\centering
\epsfig{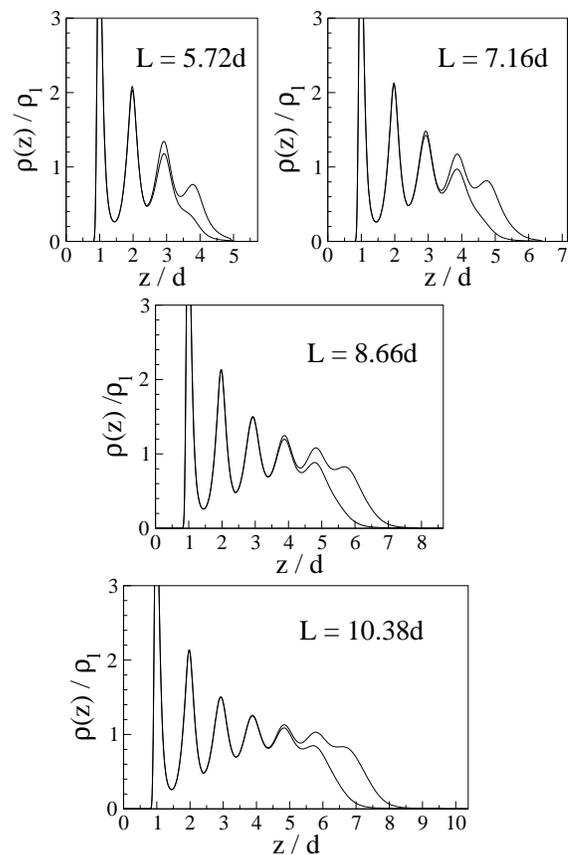}
\caption[Coexisting density profiles at the 4th, 5th, 6th and 7th layering transitions at bulk liquid--gas coexistence $\mu=\mu_{co}$ and $T=0.56T_C$ for system (i) plotted as a function of reduced chemical potential and inverse wall separation $d/L$.]{Coexisting density profiles at the 4th, 5th, 6th and 7th layering transitions at bulk liquid--gas coexistence $\mu=\mu_{co}$ and $T=0.56T_C$ for system (i) (solvophilic wall and purely repulsive solvophobic wall). $\rho_l$ is the density of the liquid at bulk coexistence; at this temperature $\rho_ld^{3}=0.79$. Note that there is also a first order transition from 0 to 1 layers for $\mu=\mu_{co}$ at $L=1.7d$ which is not shown here (see Fig.\ \ref{fig:PTWads}).}
\label{fig:PTWprofs}
\end{figure}

Figure \ref{fig:PTWprofs} displays the coexisting density profiles at the 4th, 5th, 6th and 7th layering transitions in the confined fluid. The density profiles are very similar to the coexisting density profiles next to the solvophilic wall at layering transitions in the semi-infinite system (Fig.\ \ref{fig:sixprofs}), where the transitions occur as the chemical potential approaches bulk liquid--gas coexistence from the gas side. In all the profiles the peak adjacent to the hydrophilic wall has been shown cropped because it has a very large maximum. Subsequent peaks are successively broader and lower in height and this is reflected in the difference in $L$ between layering transitions. This is greater between higher transitions. The adsorption increase is also greater for higher transitions as is seen in the heights of the steps in the adsorption isotherm (Fig.\ \ref{fig:PTWads}). Further from the solvophilic wall the oscillations in density become less pronounced and the density tails off gently from the final maximum towards a low density region, similar to that of the bulk gas, near to the hydrophobic wall. Between layering transitions increasing $L$ increases the extent of this low density region until there is enough space for another liquid layer and then a new peak appears in the density profile.

\begin{figure}
\centering
\epsfig{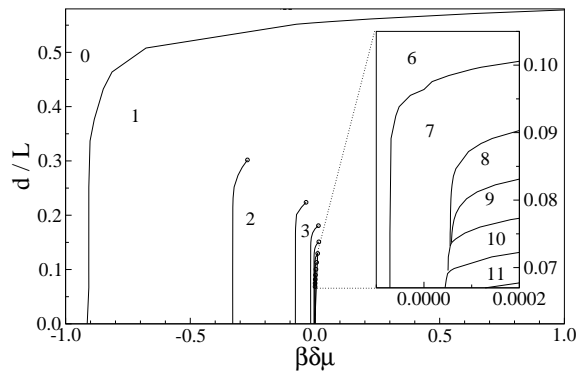}
\caption[Phase diagram showing the layering transition lines and their critical points at fixed $T=0.56T_C$ for system (i)]{Phase diagram showing the layering transition lines and their critical points $\circ$ at fixed $T=0.56T_C$ for system (i) plotted as a function of reduced chemical potential and inverse wall separation $d/L$. The inset magnifies the region close to bulk coexistence. The 7th layering transition crosses bulk coexistence $\delta\mu=0$ at $L=10.38d$ and for large wall separation $L$ lies on the bulk gas side of coexistence $\delta\mu<0$. As $L\rightarrow \infty$ the first 7 transitions occur at values of $\delta\mu$ given in Fig.\ \ref{fig:sixprofs}. The 8th and higher transition lines lie entirely on the liquid side of coexistence $\delta\mu>0$ and merge as $L\rightarrow\infty$.}
\label{fig:PTWphase}
\end{figure}

Figure \ref{fig:PTWphase} displays the equilibrium layering transition lines as functions of inverse wall separation $d/L$ and chemical potential difference from coexistence, $\delta\mu=\mu-\mu_{co}$. As the wall separation decreases the layering transitions shift to higher values of $\mu$. This is in contrast to the situation for identical parallel solvophilic walls (or in cylindrical pores \cite{BallEv}) where the shift is towards lower $\mu$ because the attractive walls favour the high density liquid phase encouraging the formation of layers when the system is more undersaturated with respect to bulk coexistence. Symmetrically confined fluids undergo capillary condensation on the approach to bulk coexistence, at which the pore completely fills with the liquid phase. There is no capillary condensation transition in our competing walls slit and the 1st, 4th, 5th, 6th and 7th layering transition lines cross bulk coexistence $\delta\mu=0$, see inset in Fig.\ \ref{fig:PTWphase}. The 2nd and 3rd layering transitions end in critical points before reaching bulk coexistence whereas the 4th--7th layering transition lines end in critical points on the liquid side of coexistence $\delta\mu>0$. Note that the 1st layering transition persists well into the bulk liquid phase, occurring at an almost constant wall separation $L\approx1.7d$ for $\beta\delta\mu>-0.5$. As $L\rightarrow\infty$ the transitions tend towards those of the semi-infinite fluid at the hydrophilic wall, i.e., there are seven transitions as bulk coexistence is approached from the gas side and these occur at the values of $\delta\mu$ given in Fig.\ \ref{fig:sixprofs}. On crossing bulk coexistence the system for $L\rightarrow \infty$ should  become filled with the liquid. When $L$ is finite there is no transition at bulk coexistence because there is still a layer of gas next to the hydrophobic wall, which would be completely dry in isolation. This means that further layering transitions are possible for $\delta\mu>0$ and we have plotted the 8th to the 11th transition lines (see inset in Fig.\ \ref{fig:PTWphase}). As these transition lines approach bulk coexistence they merge together so that for fixed large $L$ there is usually just one transition for $\delta\mu>0$ in which the number of layers jumps from seven to some large number on increasing $\mu$. As $L\rightarrow\infty$ the size of this jump in the adsorption (i.e., the number of layers) diverges corresponding to condensation of liquid (note there is still a layer of gas next to the solvophobic wall) and the chemical potential at the transition approaches bulk coexistence, $\delta\mu_{jump}\rightarrow 0^+$. 

\section{The solvation force}
\label{sec:fs}
\begin{figure}[htbp]
\centering
\epsfig{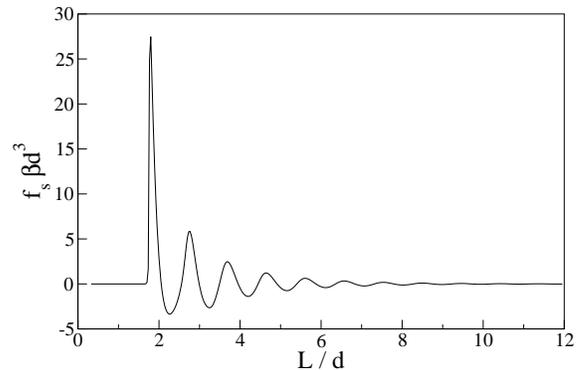}
\caption[Solvation force $f_s$ versus wall separation $L$ at $T=0.56T_C$ for the liquid confined between identical solvophilic walls]{Solvation force $f_s$ versus wall separation $L$ at bulk coexistence, $\mu=\mu_{co}$ and $T=0.56T_C$ for the liquid confined between {\em{identical}} solvophilic walls; the wall--fluid potential is given by Eq.\ (\ref{eq:wpot}).}
\label{fig:PTWfs_attID}
\end{figure}
\begin{figure}[htbp]
\centering
\epsfig{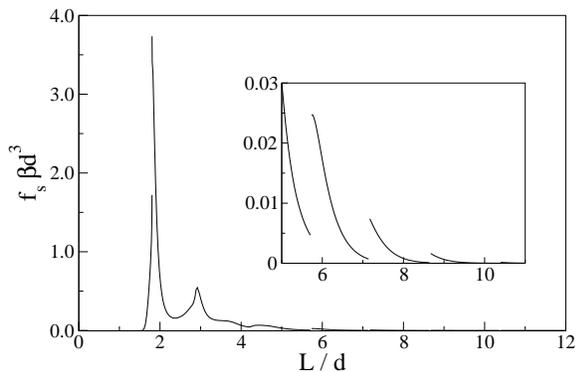}
\caption[Solvation force $f_s$ versus wall separation $L$ at $T=0.56T_C$ for the asymmetric system (i)]{Solvation force $f_s$ versus wall separation $L$ at bulk coexistence, $\mu=\mu_{co}$ and $T=0.56T_C$ for the asymmetric system (i)---a solvophilic and a purely repulsive solvophobic wall. The inset magnifies the solvation force at larger values of $L$. The jumps in $f_s$ occur at the layering transitions indicated in Figs.\ \ref{fig:PTWads} and \ref{fig:PTWprofs}.}
\label{fig:PTWfs}
\end{figure}
\begin{figure}
\centering
\epsfig{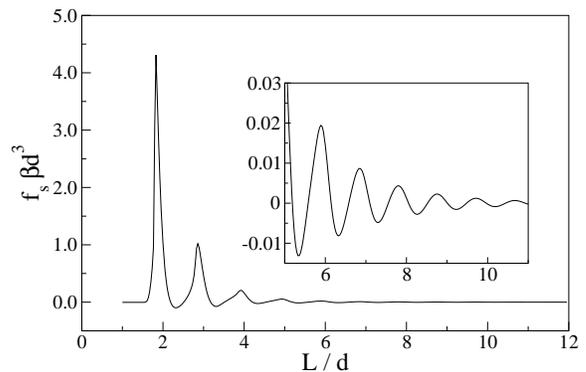}
\caption[Solvation force $f_s$ versus wall separation $L$ at $T=0.56T_C$ for the asymmetric system (ii)]{Solvation force $f_s$ versus wall separation $L$ at bulk coexistence, $\mu=\mu_{co}$ and $T=0.56T_C$ for the asymmetric system (ii)---a solvophilic and a weakly attractive solvophobic wall. The inset magnifies the solvation force at larger values of $L$. There are no jumps in $f_s$ as there are no layering transitions in this system.}
\label{fig:PTWfs_att}
\end{figure}

The solvation force (or excess pressure) $f_s$, i.e., the force per unit area between the two confining surfaces due to the intervening fluid, is plotted as a function of $L$ for three different systems in Figs.\ \ref{fig:PTWfs_attID}, \ref{fig:PTWfs} and \ref{fig:PTWfs_att}. It was calculated\cite{Evans90,StewEv12}, using DFT, from the change in equilibrium excess grand potential with $L$, 
\begin{equation}
f_s=-\frac{1}{A}\left(\frac{\partial \Omega_{ex}(L;\mu)}{\partial L}\right)_{T,\mu}.
\label{eq:fs}
\end{equation}
The solvation force of liquids confined between two identical attractive walls is well studied within non-local density functional theories \cite{BalBerGub,MacDrzBry}, in molecular simulations \cite{LanSpu,MegSno,GaoLueLan} and experimentally \cite{Isr,IsrB}. Generally for identical walls at small separations, up to around $L=10d$, the force oscillates about zero as a result of packing effects in the liquid; roughly speaking the repulsive maxima occur when the layers are compressed while attractive minima lie at wall separations where the molecules pack comfortably. Figure \ref{fig:PTWfs_attID} displays our present results for the solvation force for the liquid confined between identical solvophilic walls with the wall--fluid potential given by Eq.\ (\ref{eq:wpot}). As anticipated, the force oscillates with a period close to the molecular diameter $d$. There also appears to be a monotonically decaying repulsive component.

Our results for the solvation force for the asymmetrically confined system (i), shown in Fig.\ \ref{fig:PTWfs}, exhibit repulsive maxima just after layering transitions when there is barely sufficient space for a new layer. In contrast with the solvation force between identical walls (Fig.\ \ref{fig:PTWfs_attID}) in which the force oscillates between being positive and negative, the solvation force for the fluid confined between our attractive and repulsive walls is always positive, i.e., it is always repulsive. Between the higher layering transitions (4th and above) the solvation force decays monotonically towards zero as the wall separation is increased, see inset in Fig.\ \ref{fig:PTWfs}. This reflects the behaviour of the grand potential energy between transitions, which decreases slowly with $L$ as the amount of space occupied by fluid at the gas density increases so reducing the interaction energy between the solvophobic wall and the high density liquid layers. The shape of the solvation force versus wall separation plot at small wall separations ($L<5d$) is less easy to explain. There is a strong maximum at the 1st layering transition followed by another peak in the region of the continuous transition from one to two layers. The continuous transition from two to three layers appears to coincide with a minimum in the solvation force and this is followed by a broad maximum.

System (ii) consists of the fluid confined between the attractive, solvophilic wall and a weakly attractive but solvophobic wall (see Sec.\ \ref{sec:WFP}). The equilibrium state for this system at bulk coexistence $\mu=\mu_{co}$ is the condensed state, i.e., the slit is filled with liquid and there is no thick gas layer next to the solvophobic wall and accordingly there is no liquid--gas interface, see Fig.\ \ref{fig:PTWprofsTa} below. Consequently this system does not exhibit layering transitions on varying the wall separation $L$ at bulk coexistence $\mu=\mu_{co}$ and as a result the solvation force as a function of $L$ is continuous (Fig.\ \ref{fig:PTWfs_att}). Unlike the solvation force for system (i), which is repulsive for all wall separations $L$ (Fig.\ \ref{fig:PTWfs}), the solvation force for system (ii) (Fig.\ \ref{fig:PTWfs_att}) oscillates about zero between attraction and repulsion. These oscillations appear to be superimposed on a monotonically decaying repulsive component. The ratio of the amplitude of the oscillations to the amplitude of the monotonic component is relatively small compared to the same ratio for the solvation force between identical solvophilic walls (Fig.\ \ref{fig:PTWfs_attID}). The oscillations in the solvation force at small wall separations $L$ are much less regular for system (ii) than for the symmetric solvophilic system. At larger wall separations, $L\gtrapprox5d$, the oscillations are quite uniform in both of these systems (see inset to Fig.\ \ref{fig:PTWfs_att}). Note that overall the solvation force in the asymmetrically confined system is much weaker than in the symmetrically confined system at the same temperature and chemical potential.

\section{Comparison with the Results of Pertsin and Grunze}
\label{sec:PG}
In this section we compare results from DFT with those from grand canonical Monte Carlo simulations of water confined in an asymmetric slit, performed by Pertsin and Grunze (PG) \cite{PertGrun}, using the TIP4P model for water \cite{JorgChand}. We used essentially the same hydrophilic wall--water potential as the unstructured hydrophilic wall of PG and two different hydrophobic walls---(i) a purely repulsive wall, and (ii) a weakly attractive wall (potentials given in Sec.\ \ref{sec:Model}). In the results below we fixed the wall separation at $L=10d=30$\AA.

\begin{figure}
\centering
\epsfig{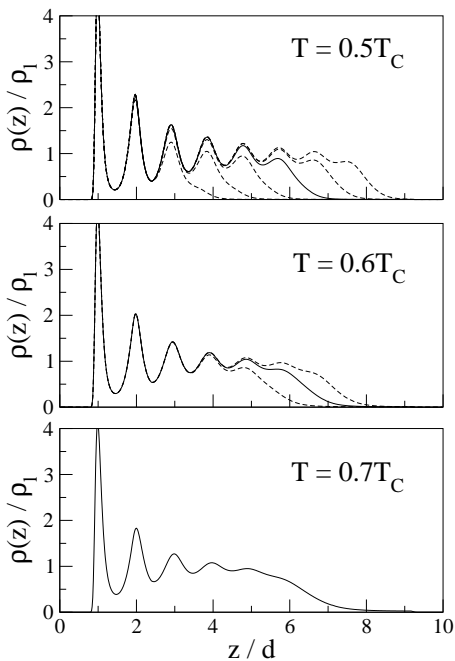}
\caption[Density profiles for $L=10d$ for three temperatures at bulk coexistence, $\mu_{co}$, for system (i)]{Density profiles for $L=10d$ for three temperatures at bulk coexistence, $\mu_{co}$, for system (i). The liquid densities at bulk coexistence are $\rho_ld^{3}=0.84$, $0.75$ and $0.66$ at $T/T_C=0.5$, $0.6$ and $0.7$ respectively. The left-hand wall is solvophilic and right-hand wall is solvophobic and purely repulsive (i). In each case the full line is the equilibrium state whilst the dashed lines are metastable states. Note the presence of a thick layer of gas near the right-hand solvophobic wall for the equilibrium profiles.}
\label{fig:PTWprofsT}
\end{figure}

\begin{figure}
\centering
\psfig{figure=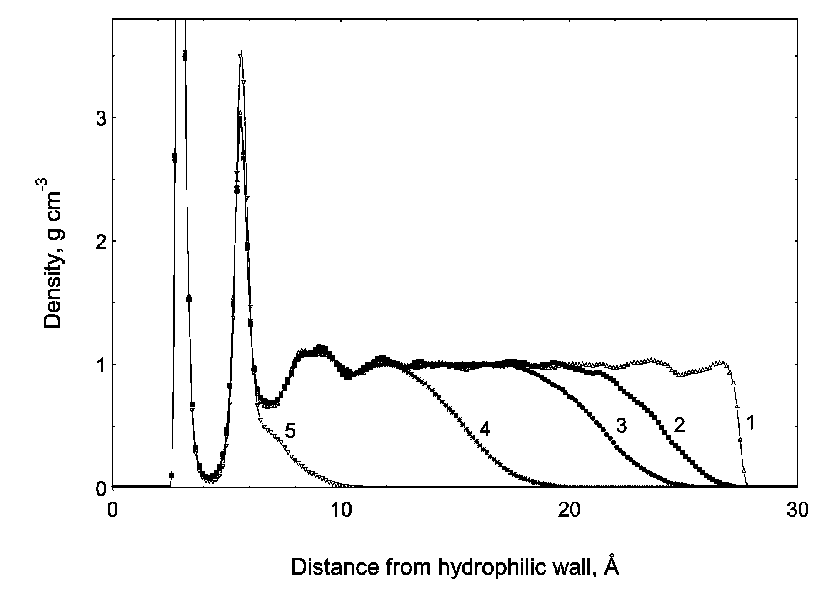, width=8.5cm, height=6cm}
\caption[Density profiles of water obtained by Pertsin and Grunze using grand canonical Monte Carlo simulations (adapted, with permission, from Fig.\ 3 Ref.\ \citenum{PertGrun}. Copyright 2004 American Chemical Society.)]{Density profiles of water obtained by Pertsin and Grunze using grand canonical Monte Carlo simulations (adapted, with permission, from Fig.\ 3 Ref.\ \citenum{PertGrun}. Copyright 2004 American Chemical Society.). The wall separation was $L=30$\AA, i.e., roughly 10 molecular diameters and the chemical potential was $\mu=-6.105$ kcal/mol (bulk coexistence was at $\mu_{co}=-6.2 \pm 0.05$ kcal/mol). Curve 1 is the density profile for fluid in a slit with the same wall--fluid potentials as our system (ii), a weakly attractive solvophobic wall. The slit is filled with liquid. Curves 2-5 are for fluid in a slit with the same wall--fluid potentials as our system (i) (repulsive solvophobic wall); curve 5 corresponds to a metastable state. The surface tensions (excess grand potentials) of the density profiles labeled 2-4 are all very similar. The first density maximum is cropped.}
\label{fig:PTWprofPG}
\end{figure}

Figure \ref{fig:PTWprofsT} shows density profiles obtained from our DFT model for system (i) at three different temperatures, all at bulk liquid--gas coexistence $\mu=\mu_{co}$. The simulations of PG were performed at room temperature, which corresponds to $\approx0.49T_C$ in the TIP4P model \cite{NezKol}. The exact value of the coexisting chemical potential in the simulations was not known but according to an estimate by the authors the state point of their investigations was very near to and on the liquid side of bulk coexistence---see caption to our Fig.\ \ref{fig:PTWprofPG}. Our results should be compared with those of PG, displayed in Figure\ \ref{fig:PTWprofPG}, where curve 5 is a metastable state with two high density liquid layers next to the attractive wall and curves 2-4 have seven, six and four layers respectively and have surface tensions (excess grand potentials) which are approximately equal, within the statistical uncertainty of the simulations. We did not find a metastable state with two liquid layers at any of the three temperatures but at $T=0.5T_C$ we observed minima in the excess grand potential corresponding to states with 3 to 8 liquid layers. The equilibrium state had 6 layers but the profiles with 5-8 layers all had very similar excess grand potentials---see Fig.\  \ref{fig:PTWprofsT}. Comparing the shape of the density profiles in Fig.\ \ref{fig:PTWprofsT} for $T=0.5T_C$ with those of PG for water we observe that the heights of the peaks of the first two liquid layers are much greater in the PG simulation results and the subsequent layers show less pronounced oscillations than our profiles. The shape of the PG profiles after the first two peaks is in fact closer to our results at higher temperatures, $T=0.6T_C$ and $T=0.7T_C$; however at $T=0.6T_C$ we only found three minima in the grand potential at 5, 6 and 7 layers (Fig.\ \ref{fig:PTWprofsT}) and at $T=0.7T_C$ there were no metastable states. The first two peaks in the simulation density profiles in Fig.\ \ref{fig:PTWprofPG} may be enhanced by hydrogen bonding between water molecules, which is possible in the TIP4P model. This is not present in our much simpler model fluid. The shape we obtain for the decay of the density into the gas region near to the repulsive wall is in good agreement with the results of PG.

\begin{figure}
\centering
\epsfig{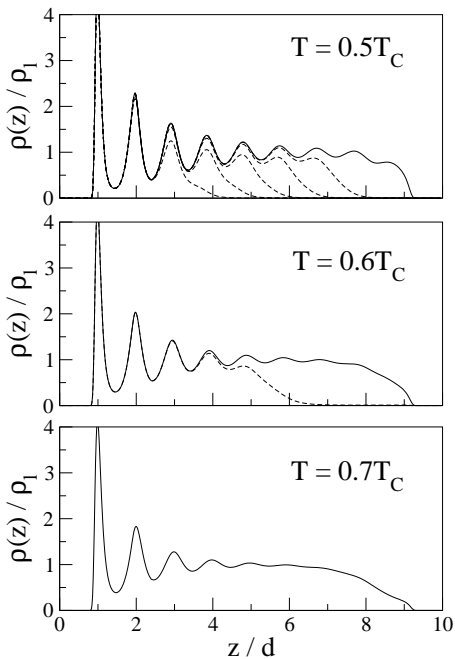}
\caption[Density profiles for $L=10d$ for three temperatures at bulk coexistence for system (ii)]{Density profiles for $L=10d$ for three temperatures at bulk coexistence, $\mu_{co}$, for system (ii). The liquid densities $\rho_l$ are given in Fig.\ \ref{fig:PTWprofsT}. The right-hand wall potential includes a weak attractive part (ii). In each case the full line is the equilibrium state which corresponds to the slit pore filled with liquid. The dashed lines are metastable states; these exhibit thick layers of gas.}
\label{fig:PTWprofsTa}
\end{figure}

When the solvophobic wall is made weakly attractive (see system (ii) Sec.\ \ref{sec:Model}) it is no longer completely dry at bulk coexistence at any of the three temperatures $T=0.5T_C$, $T=0.6T_C$ and $T=0.7T_C$; the corresponding contact angles are calculated to be $159^{\rm o}$,  $162^{\rm o}$ and  $168^{\rm o}$ respectively \cite{StewThesis}. The density profiles for the {\em semi-infinite liquid} adsorbed at the isolated wall exhibit no gas layer intruding between the wall and the liquid but instead a small region of reduced density which is of greater extent at higher temperatures $T$. When the fluid is confined between the solvophilic wall and this weakly attractive wall (system (ii))  in the equilibrium state the pore becomes filled with liquid (Fig.\ \ref{fig:PTWprofsTa}). The density profile for the equilibrium state at $T=0.5T_C$ (top panel Fig.\ \ref{fig:PTWprofsTa}) is the closest to that of PG for the same confining walls (curve 1 Fig.\ \ref{fig:PTWprofPG}). Our results show a slight decrease in density in the final layer of liquid at the solvophobic wall near $z=8.7d$ whereas in the PG profiles this maximum has a similar height to those near to the centre of the slit. In addition to the liquid filled equilibrium states we observe {\em{metastable}} states at $T=0.5T_C$ and $T=0.6T_C$ (top and middle panels Fig.\ \ref{fig:PTWprofsTa}) in which there is a film of gas phase near to the hydrophobic wall, similar to the states seen in system (i) (Fig.\ \ref{fig:PTWprofsT}). The corresponding minima in the excess grand potential for these metastable states are shallow and significantly higher in free energy than the global minimum of the liquid filled state, which may explain why no such metastable states were observed in the PG simulations.

\section{Summary and Discussion}
\label{sec:Dis}
Using classical DFT we have observed a series of first order transitions as successive layers of a Lennard-Jones like fluid condense on a single solvophilic substrate as liquid--gas coexistence is approached from the gas side. We also investigated the properties of the same fluid confined between a solvophilic and a solvophobic wall. In our first system, (i), in which the wall--fluid potential for the solvophobic wall was purely repulsive, layering transitions were found at chemical potentials shifted from the corresponding values in the semi-infinite fluid. Layering transitions were {\em{not}} found in system (ii), which has a weakly attractive solvophobic wall. In system (i) the solvation force (i.e. the excess pressure resulting from confinement) was repulsive at all wall separations $L$ and was discontinuous at each first-order layering transition whereas for system (ii) the solvation force oscillated between attraction and repulsion but with a monotonically decaying repulsive component. DFT results for our simple model fluid confined in an asymmetric slit accounted for the main features of grand canonical Monte Carlo simulations\cite{PertGrun} of a realistic model of water for $\mu$ close to $\mu_{co}$; we found similar density profiles.

The pronounced heights of the first few peaks and the low values at the minima of the density profiles (Figs.\ \ref{fig:sixprofs}, \ref{fig:PTWprofs} and \ref{fig:PTWprofsT}) raise the possibility that the layers closest to the solvophilic wall may be frozen. Indeed, very rich phase behaviour involving solid-like adsorbed films has been observed experimentally. For example the condensation of further fluid layers of oxygen on graphite causes the innermost layers to freeze \cite{YounHess} and solid argon adsorption on graphite exhibits re-entrant layering, i.e., the layering steps in the adsorption isotherm become broad as temperature is increased but reappear at higher temperatures \cite{ZhuDash,ZhuDashB,YounMengHess}. The bulk triple point temperature for a Lennard-Jones fluid depends on how the pair potential is truncated but from simulations \cite{FanMon,AhmSad} this is known to be $T_{tr}\lesssim 0.55T_C$ . Below and slightly above the bulk triple point temperature, the adsorbed layers closest to the wall may be crystalline. Were our fluid to freeze we would have to allow for variations in the density profile parallel to the wall as well as perpendicular to it, i.e. we would need to solve for $\rho(x,y,z)$ rather than just $\rho(z)$. Since our present implementation of DFT does not allow for freezing we must turn to simulation to obtain estimates for the temperatures at which the layers will freeze. Results for molecular simulations of (truncated) LJ fluids in contact with attractive substrates are found in Refs.\ \citenum{RadGubSli,SchQuiHen}. Both of these simulations used a ``10-4-3'' Steele potential \cite{Steele} for the wall--fluid interaction in contrast to our 9-3 potential, Eq.\ (\ref{eq:wpot}). In order to relate these results to our model, we employ the dimensionless ratio $\epsilon_r$, introduced in \cite{SchQuiHen}, which is obtained by dividing the minimum of the wall--fluid potential by $\epsilon$, the strength of the fluid--fluid interaction. The minimum of the potential given in Eq.\ (\ref{eq:wpot}) occurs at $z_{min}=\zeta$ and has depth $-V_w(z_{min})=6.2$kcal/mol. The ratio for our system is therefore $\epsilon_r=4$. In Ref.\ \citenum{RadGubSli}, for wall strengths corresponding to $4<\epsilon_r<12$, the authors find (using molecular simulations) that the layer of fluid in contact with the wall crystallises at temperatures varying from $T_b$ to $1.2T_b$, where $T_b$ is the bulk freezing temperature and the results in Ref.\ \citenum{SchQuiHen} are consistent with this. Thus provided we confine our attention to temperatures above the bulk triple point then the layering transitions that we have identified are likely to be between different {\em{liquid-like}} adsorbed phases. That the density profiles, for all the state points that we consider here, do not exhibit extremely low values at the minima gives us further confidence that we are considering fluid phases.

Close to the triple point temperature, our DFT results show six high density layers adsorbed at a single solvophilic wall at bulk coexistence and the number of layers increases discretely in a number of layering transitions as temperature is increased along the line of bulk coexistence (Fig.\ \ref{fig:PhaseD}). One might ask what results our model would yield as the temperature is increased from the upper limit of our investigation, about
$0.65T_C$, towards the bulk critical temperature. Provided the grand potential $\Omega(\Gamma_{ex})$ exhibits oscillatory decay, e.g. Fig.\ \ref{fig:ExGP0_56}, then its minimum value occurs for a finite adsorption $\Gamma_{ex}$ and the wall will not be completely wet \cite{Hend94}. The oscillatory behaviour of the grand potential arises from the binding potential term $\omega(l)$, which is the interaction energy between the wall--liquid and liquid--gas interfaces as a function of their separation, $l$; for large separations $l\propto\Gamma_{ex}/A$. In the
absence of any long-ranged intermolecular forces the binding potential
originates from the interaction between the tails of the wall--liquid and liquid--gas density profiles. For finite-ranged intermolecular potentials of the type we consider here the asymptotic decay of the density profiles has the same functional form as the decay of the bulk radial distribution function $g(r)$ \cite{EvHendHoy,EvLdC}. Thus from general considerations\cite{Hend94,Hend05} one expects the asymptotic decay of the grand potential to be monotonic at high temperatures and complete wetting to occur. At some intermediate temperature $T_{FW}$, where the Fisher-Widom line meets the bulk coexistence curve, crossover occurs and it is near this temperature that we expect a wetting transition to occur in the present mean-field DFT treatment\cite{Hend94,Hend05}. We do not speculate here on whether complete wetting would occur via an infinite series of layering transitions or via some other scenario---see below. Rather we note that in an experimental situation long-range dispersion interactions, with power-law decaying potentials, will always be present. These forces always dominate the ultimate decay of the density profiles and therefore of the interfacial binding potential and grand potential, thus ruling out the possibility of wetting via an infinite sequence of layering transitions. The lower order layering transitions will remain but eventually complete wetting must occur, driven by the power-law attraction, and this might occur at a temperature below $T_{FW}$, i.e. below that predicted by a treatment based on the corresponding short-ranged (truncated) potential description.

We find that the critical point temperatures for (at least) the first four layering transitions, $T_{cn}$ increase with $n$ (Fig.\ \ref{fig:PhaseD}). The temperature $T_{FW}$ is relevant here. Above this temperature the asymptotic decay of the interfacial binding potential is monotonic so in this region of the phase diagram there cannot be an infinite series of layering transitions. Lower order transitions may still be possible for $T>T_{FW}$ but it seems more likely that $T_{cn}<T_{FW}$ for all the layering transitions.
Our analysis up to this point has been purely mean-field and we have ignored fluctuation effects. We expect that thermal fluctuations would damp the oscillations in the density profiles and therefore reduce the amplitude of the oscillations in the grand potential as a function of adsorption. Higher order transitions may be completely washed out or may merge into a single prewetting transition. However, as mentioned in Section \ref{sec:int}, experimentally more than eight layering transitions have been observed for certain molecular fluids\cite{DrirNhamHess,DrirHess} adsorbed on graphite at temperatures above $T_{tr}$ and we believe that for our model system at least the first few layering transitions should persist beyond mean-field.
Of course our results for layering transitions apply to idealized smooth surfaces. In a study of adsorption of nitrogen and argon in carbon pores, Neimark {\em et. al.} \cite{NeiLin} used Quenched Solid Density Functional Theory (QSDGFT), which allows for some inhomogeneity in the surface, to show that step wise isotherms become smooth and sharp layering transitions are no longer present once the surface is made rough.

We conclude this discussion of layering transitions at a single wall by comparing our results to previous work in this area. Our density profiles (Fig.\ \ref{fig:sixprofs}) and adsorption isotherms (Fig.\ \ref{fig:Ads0_56}) for the semi-infinite fluid at a solvophilic wall resemble those obtained by Ball and Evans \cite{BallEv} from a weighted density DFT approximation. Those authors could only speculate on a surface phase diagram for their system. Here we have determined the equilibrium density profile and grand potential for a very large number of state points, allowing us to trace accurately three layering transition lines to their critical points (Fig.\ \ref{fig:PhaseD}). Our phase diagram is in qualitative agreement with mean-field {\em lattice gas} results for an intermediate strength substrate (see Figs.\ 13 and 17a in Ref.\ \citenum{PanSchWor}). 
DFT and simulation results for a model colloid-polymer mixture at a hard-wall \cite{BradEvSch,DijVR,WesSchLow} show several layering transitions (depending on the choice of parameters) along the colloid gas side of two-phase coexistence followed by a first order wetting transition. In contrast to our present results the first few layering transitions occurring in these colloid-polymer mixtures are far from the bulk triple point. Consequently the coexisting (colloid) density profiles at the layering transitions are less highly structured than for the present simple fluid (Fig.\ \ref{fig:sixprofs}).

Turning now to the {\em confined} fluid we focus first on the results for the solvation force which can be investigated experimentally using the surface force apparatus \cite{Isr,IsrB,ButCapKap}. In the latter the force between two surfaces, separated by a distance $L$, is measured in crossed cylinder geometry. This quantity is easily related to the solvation force $f_s$ between planar surfaces using the well-known Derjaguin approximation \cite{Der1934}. The surfaces are usually molecularly smooth mica which may be coated by a film of some other material. The solvation force measured between identical substrates is usually oscillatory in $L$ at small separations ($L<10$ molecular diameters), reflecting the local ordering of the fluid arising from packing effects in confinement. Our DFT results for the fluid confined between two identical solvophilic walls yield a damped oscillatory solvation force (Fig.\ \ref{fig:PTWfs_attID}) in general agreement with experimental results for simple inorganic liquids \cite{ChrisClaes,HorIsr} confined between two mica surfaces and with many other theoretical and simulation studies of simple model fluids.
There have been very few systematic experimental studies of the solvation force for a fluid between different surfaces \cite{ChrisClaes,KokZuk}. In our theoretical study we focus on the situation where one wall is solvophilic and the other is solvophobic. For both our asymmetric systems at bulk coexistence $\mu=\mu_{co}$ and $T=0.56T_C$ we found that the solvation force had a repulsive monotonic component (Figs.\ \ref{fig:PTWfs} and \ref{fig:PTWfs_att}). The solvation force for system (i) (see Fig.\ \ref{fig:PTWfs}), is repulsive at all wall separations $L$, and is discontinuous at each first order layering transition, taking a maximal value immediately after a transition. In system (ii) the solvophobic wall is weakly attractive; this represents a more realistic physical situation as the corresponding contact angle $\theta$ is large but $<\pi$. The solvation force for system (ii) oscillates between repulsion and attraction but still has a strong monotonically decaying repulsive part (Fig.\ \ref{fig:PTWfs_att}). Repulsion has been observed experimentally for NaCl solutions between a hydrophobic sphere and a hydrophilic surface but in this case the surface was charged and the force was most likely electrostatic in origin \cite{KokZuk}. We are not aware of any experimental observations of oscillatory $L$ dependence for the solvation force as is predicted by our model (Figs.\ \ref{fig:PTWfs} and \ref{fig:PTWfs_att}). For these oscillations to be seen experimentally the two surfaces would have to be smooth and quite close together ($L\lessapprox10d$).

Theoretical and simulation studies of fluids confined by non-identical walls are also rather scarce. Parry and Evans \cite{ParryEvPRL,ParryEv} investigated the behaviour of the simplest (Landau) model of a fluid confined between a wetting and a drying wall focusing on the localization-delocalization phase transition for large $L$. The study of Stewart and Evans \cite{StewEv12} used the same DFT approach as the present but concentrated on the properties of the high temperature ($T>T_W$) delocalized interface phase at large $L$. These authors also compared their results with those of Monte Carlo simulations of an asymmetrically confined Asakura-Oosawa-Vrij model of a colloid-polymer mixture\cite{VirVinHorBinPRE}. Earlier Balbuena {\em{et. al.}} \cite{BalBerGub} used a nonlocal DFT to study a truncated Lennard-Jones fluid confined between two walls exerting (10-4-3) potentials of different strengths for wall separations of $L<9\sigma$. Their results \cite{BalBerGub}, which correspond to a low density {\em bulk} gas, show capillary condensation (pore filling) with the `liquid' exhibiting highly structured density profiles and a solvation force which oscillates as $L$ is reduced---similar to their findings for identical solvophilic walls. No layering transitions were observed at the fairly high temperatures which the authors investigated. The effect of reducing the strength of the attractive wall--fluid potential at one of the walls was to shift the separation at which condensation occurs to lower values and a decrease in the amplitude of the oscillations (see Fig.\ 8a in Balbuena {\em{et. al.}} \cite{BalBerGub}). By contrast we find the solvation force in asymmetric slits (Figs.\ \ref{fig:PTWfs} and \ref{fig:PTWfs_att}) displays significant monotonic repulsive parts in addition to an oscillatory component and for system (i) is repulsive at all wall separations $L$. However the wall potentials used by Balbuena {\em{et. al.}} \cite{BalBerGub} are more strongly attractive than those considered in this paper---the least solvophilic wall potential has $\epsilon_r=2.15$ compared to $\epsilon_r=0.3$ for our weakly attractive but solvophobic wall (where $\epsilon_r$ is the ratio of the minimum in the wall potential $-V_w(z_{min})$ to the fluid--fluid potential strength $\epsilon$). Whilst our density profiles exhibit a low density region next to the solvophobic wall, which for the purely repulsive potential (i) takes values as low as the coexisting gas density, the profiles in Fig.\ 8a in Balbuena {\em{et. al.}} \cite{BalBerGub} have pronounced peaks even for the least attractive walls. Having a film of gas next to the repulsive solvophobic wall may be necessary to obtain a solvation force which is always repulsive. 

Although we did not aim to account for the detailed behaviour of confined water (clearly our Lennard-Jones model cannot capture features arising from hydrogen bonding) we were influenced by some of the phenomena that have been observed in simulation and experimental studies of asymmetrically confined water. We comment on the most relevant.
In Section \ref{sec:PG} we compared density profiles obtained from our DFT study with those found by Pertsin and Grunze (PG) \cite{PertGrun} using Monte Carlo simulations for TIP4P water (Fig.\ \ref{fig:PTWprofPG}). Indeed the striking results of PG\cite{PertGrun} provided some impetus for our investigation. Our results for system (i), in which the solvophobic wall was purely repulsive, are similar to those of PG\cite{PertGrun}. We found several metastable states with very similar grand potentials corresponding to different numbers of layers of liquid and our density profiles (Fig.\ \ref{fig:PTWprofsT}) showed a layer of fluid with the gas density next to the solvophobic wall. Leaving aside the first two layers adsorbed at the solvophilic wall, our density profiles show more pronounced structure than that found in the simulations for the same reduced temperatures. It is likely that our DFT exaggerates the oscillations in the density profile. Fan and Monson \cite{FanMon} compared density profiles obtained using Monte Carlo simulations of a truncated Lennard-Jones ($12,6$) fluid at a single Lennard-Jones ($9,3$) wall with those obtained using a weighted density approximation for the hard-sphere free energy functional in a DFT approach and found that the DFT overestimated the amplitude of the oscillations in the density profile, especially at low temperatures ($T \approx T_{tr}$).

In the second system (ii) that we investigated in Section \ref{sec:PG}, the repulsive solvophobic wall was replaced by a weakly attractive substrate. The contact angle for the fluid at this wall was $\theta\approx160^{\rm o}$ depending on the temperature, making this wall more akin to the type of hydrophobic surface likely to be seen experimentally than the completely dry wall of system (i). Once again our results were in good overall agreement with the simulations of PG \cite{PertGrun}. Equilibrium density profiles (solid lines in Fig.\ \ref{fig:PTWprofsTa}) showed that the slit was completely filled with liquid with no low density gas layer near to the solvophobic wall. We conclude that whilst our simple model cannot account for specific details of PG’s water simulations\cite{PertGrun}, such as the enhanced density of the first two adsorbed layers at the hydrophilic wall, it does appear to capture the main features.

PG’s study\cite{PertGrun} was prompted in part by SFA measurements of Zhang {\it et. al.} \cite{ZhaZhuGran} for water confined between a hydrophobic and a hydrophilic surface. Their results, which correspond to a bulk reservoir of water at normal temperature and pressure, showed very large fluctuations in dynamical response to shear deformations and the authors conjectured these were due to the presence of a fluctuating liquid--gas interface. They termed their system a Janus interface. In our earlier paper\cite{StewEv12} we argued that it was difficult to see why the particular choice of wet (mica) and {\em partially} dry (mica coated by hydrophobic layers) walls studied in the SFA experiments \cite{ZhaZhuGran} should give rise to a wildly fluctuating interface of the type associated with the delocalized soft-mode phase studied by Parry and Evans \cite{ParryEv} and in Ref.\ \citenum{StewEv12}. We return to the argument in the light of our present study which pertains more closely to the SFA experiments. First: In the SFA experiments the largest contact angle of water with the hydrophobic surface was $\theta=120^{\rm o}$ for a thiol coated surface. This situation is more akin to our weakly attractive solvophobic wall system (ii), which does not exhibit a liquid--gas interface, than to our repulsive drying wall system (i), which does exhibit a liquid--gas interface. Second: Measurements of the static normal force between the two surfaces indicated attraction at all surface separations down to around 5--20 molecular (water) diameters where the surfaces sprang into contact. This is in contrast with our results for system (ii), in which the solvation force oscillated between attraction and repulsion as the surface separation $L$ increased and also had a monotonically decaying repulsive component (Fig.\ \ref{fig:PTWfs_att}). Moreover in system (i), see Fig.\ \ref{fig:PTWfs}, the solvation force was found to be repulsive at all wall separations. Note that our previous study\cite{StewEv12} showed that the presence of a fluctuating liquid gas interface, at $T>T_W$, also yields a repulsive solvation force; in this case the long-ranged dispersion interactions result in $f_s$ decaying as $L^{-3}$ for large $L$. Our results do not support the hypothesis in Ref.\ \citenum{ZhaZhuGran} that the noisy shear response they observed was due to a fluctuating liquid--gas interface and, like PG\cite{PertGrun}, we conclude that another explanation of the SFA results must be sought.

Shortly after the publication of the article by Zhang {\it et. al.} \cite{ZhaZhuGran}, Lin and Granick\cite{LinGran} explained that many of the SFA experiments in Granick's laboratory might have been influenced by the presence of Pt nanoparticles resulting from cleaving the mica; see also De Gennes\cite{DeG}.

Also prompted by the results of Zhang {\it et. al.} \cite{ZhaZhuGran}, McCormick \cite{McC} investigated a simple lattice gas model for `water' confined between hydrophilic and hydrophobic walls in which the hydrophobic wall had a contact angle of $\theta=143^{\rm o}$. A liquid--gas interface position was defined to be located at the average of the distances of the nearest liquid cell and furthest gas cell from the hydrophobic wall. The time and space averaged position of this interface was at $2d$, suggesting one or two layers of gas next to the hydrophobic wall. Large fluctuations in the location of this interface were observed which McCormick argued provided evidence for a fluctuating liquid--gas interface. This result appears to be in contrast with our results for system (ii), a slightly more hydrophobic wall ($\theta\approx160^{\rm o}$) where equilibrium density profiles (Fig.\ \ref{fig:PTWprofsTa}) showed layers of liquid next to the hydrophobic wall and therefore no liquid--gas interface. Thus there appears to be some inconsistency between our results and those obtained for the lattice gas \cite{McC}. This may be related to the manner in which the liquid--gas interface was defined in the simulations by McCormick\cite{McC}.

The physical situation that should be relevant for SFA measurements on water between hydrophilic and hydrophobic surfaces is that described by PG’s density profile\cite{PertGrun} labelled 1 in our Fig.\ \ref{fig:PTWprofPG} and the equilibrium profiles at $T=0.5T_C$ or $0.6T_C$, in Fig.\ \ref{fig:PTWprofsTa}. In the light of PG's observation\cite{PertGrun} of a large root mean square fluctuation in the number of molecules and McCormick's observation\cite{McC} of strong fluctuations in the lattice gas occupancy at contact it is tempting to speculate that the local susceptibility (or local compressibility) $\left(\frac{\partial \rho(z)}{\partial\mu}\right)_{L,T}$, as considered, for example, in Ref.\ \citenum{StewEv12}, is large for $z$ in the vicinity of the hydrophobic or solvophobic wall even when that wall is only partially dry, i.e. the contact angle is large but $<180^{\rm o}$. Indeed there is growing evidence, from simulations of a variety of water models, that water near such a hydrophobic substrate exhibits enhanced density fluctuations. Recent papers by Acharya {\it et.al.}\cite{AchHarVemSri} and Mittal and Hummer \cite{MitHum} provide a comprehensive set of relevant references. We shall return to these issues in a future publication.

Finally we return to case (i) where the solvophobic wall is purely repulsive and therefore dry. Fixing chemical potential $\mu$ and varying wall separation $L$ leads to a similar sequence of layering transitions as setting $L=\infty$ and varying $\mu$. The physical explanation for this observation lies in the fact that a layer of gas develops at the solvophobic wall and this appears to mimic the role of the gas reservoir (with $\mu<\mu_{co}$) in the semi-infinite solvophilic system. The presence of the gas layer permits the development of a gas--liquid interface in the confined fluid---see Fig.\ \ref{fig:PTWprofs}. It would be interesting to establish more formal connections between the two physical situations.

\section{Acknowledgements}
We thank R. Roth for helpful comments on the manuscript and advice on numerical methods in the early stage of this work and A.Pertsin for helpful correspondence. MCS is grateful to EPSRC for financial support during her PhD studies. R.E. acknowledges support from the Leverhulme Trust under award EM/2011-080.

\appendix
\section{Finding minima of the grand potential functional.}
\label{sec:num}
\begin{figure}
\centering
\epsfig{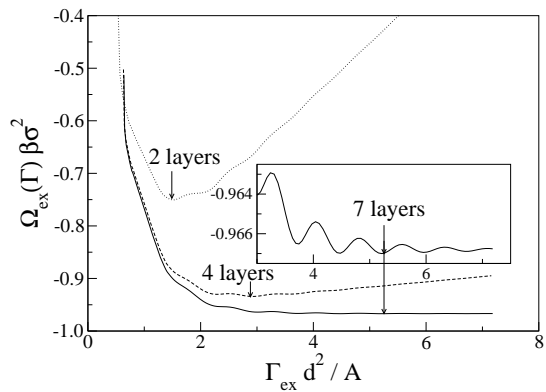}
\caption[Excess grand potential for non-equilibrium density profiles $\tilde{\rho}(z)$, as a function of adsorption for $T=0.56T_C$ at three chemical potentials]{Excess grand potential for non-equilibrium density profiles $\tilde{\rho}(z)$ adsorbed at a single solvophilic wall, as a function of excess adsorption, for $T=0.56T_C$ at three reduced chemical potentials: $\beta\delta\mu=0^-$ (full line), $\beta\delta\mu=-0.01$ (dashed line) and $\beta\delta\mu=-0.1$ (dotted line). The global minima are marked with arrows. The inset shows that the low amplitude oscillations are present in the excess grand potential for large adsorptions at bulk coexistence, $\delta\mu=0^-$.}
\label{fig:ExGP0_56}
\end{figure}

Picard iteration (details of the numerical methods used can be found in Chapter 3 of Ref.\ \citenum{StewThesis}) is usually a robust and reliable means of minimizing $\Omega_{V}[\rho]$. However, this was slow to converge because of the oscillatory nature of the density profile close to the solvophilic wall. There was also the additional problem of the large number of metastable states associated with different numbers of layers of liquid. In order to overcome these difficulties, the grand potential $\Omega_V[ \tilde{\rho}]$ was first obtained for a large number of non-equilibrium density profiles $\tilde{\rho}(z)$. These density profiles were the result of allowing a fixed number of iterations (e.g. $j=2000$) from different initial profiles. The first starting profile ($j=0$) was chosen to have either i) constant density equal to that of the bulk gas $\rho_b(T,\mu)$ or ii) a `sharp-kink' profile with density equal to that of the liquid at bulk coexistence, $\rho_l(T,\mu_{co})$, between $z=0$ and $z=l$ and that of the bulk gas for $z>l$. Subsequent starting profiles were created from the previous final profile by adding a thin layer, $\delta l$, (e.g. $\delta l = \sigma/12$) of liquid density at $z=(j-1)\delta l+l$. This method was found to be more satisfactory than using a different `sharp-kink' starting profile each time because it took advantage of the presence of peaks in the density profile near to the wall that had already formed so that the correct shape profile was reached in fewer iterations. Once the shape of the profile was correct the liquid--gas interface moved only very slowly towards a stable or metastable position. The grand potential $\Omega_V[ \tilde{\rho}] $ could then be plotted (Fig.\ \ref{fig:ExGP0_56}) as a function of the excess adsorption of the resulting density profiles:
\begin{equation}
\frac{ \Gamma_{ex}[ \tilde{\rho};T,\mu]}{A}=\int{dz [\tilde{\rho}(z;T,\mu)-\rho_b(T,\mu)]}.
\label{eq:GibbsAbs}
\end{equation}
The excess adsorption per unit area, $ \Gamma_{ex}[ \tilde{\rho};T,\mu]/A$,  provides a measure of the distance of the liquid--gas interface from the wall $l\sim \Gamma_{ex}/A(\rho_l-\rho_g)$. The minima in the grand potential $\Omega_V[\tilde{\rho}]$ correspond to positions of the liquid--gas interface with different numbers of layers of liquid. Accurate values for the grand potential at its minimum points could then be found by allowing the appropriate profiles to converge fully and layering transitions were identified by comparing the grand potential energy of the density profiles at the different minima. The global minimum at a given chemical potential corresponds to the equilibrium state and a layering transition occurs when there are two equal minima, see Fig.\ \ref{fig:ExGP0_56} for adsorption at a single solvophilic wall.

\end{document}